\documentclass{Interspeech}

\interspeechcameraready

\title{From Weak Labels to Strong Results:
Utilizing 5,000 Hours of Noisy Classroom Transcripts with Minimal Accurate Data}

\author[affiliation={1}]{Ahmed Adel}{Attia}
\author[affiliation={2}]{Dorottya }{Demszky}
\author[affiliation={3}]{Jing}{Liu}
\author[affiliation={1}]{Carol }{Espy-Wilson}

\affiliation{Electrical and Computer Engineering}{University of Maryland}{}
\affiliation{Graduate School of Education}{Stanford University}{}
\affiliation{College of Education}{University of Maryland}{}
\email{aadel@umd.edu, ddemszky@stanford.edu, jliu28@umd.edu, espy@umd.edu}
\keywords{ASR, weakly supervised learning}

\usepackage{comment}
\usepackage{multirow}
\usepackage[most]{tcolorbox}

\begin{document}

\maketitle

% the abstract here must exactly match the abstract entered into the paper submission system
\begin{abstract}
    
    % 1000 characters. ASCII characters only. No citations.
Recent progress in speech recognition has relied on models trained on vast amounts of labeled data. However, classroom Automatic Speech Recognition (ASR) faces the real-world challenge of abundant weak transcripts paired with only a small amount of accurate, gold-standard data. In such low-resource settings, high transcription costs make re-transcription impractical. To address this, we ask: what is the best approach when abundant inexpensive weak transcripts coexist with limited gold-standard data, as is the case for classroom speech data? We propose Weakly Supervised Pretraining (WSP), a two-step process where models are first pretrained on weak transcripts in a supervised manner, and then fine-tuned on accurate data. Our results, based on both synthetic and real weak transcripts, show that WSP outperforms alternative methods, establishing it as an effective training methodology for low-resource ASR in real-world scenarios.
% To better understand the impact of transcription errors, we first conduct controlled experiments using synthetically corrupted transcriptions of the TEDLIUM dataset. We then extend our analysis to a case study on classroom speech, a domain where ASR remains a challenging open problem due to data scarcity. 
% Our findings indicate that even highly corrupted transcriptions can significantly enhance ASR performance when used in conjunction with a small amount of gold-standard data, highlighting the potential of weakly supervised learning for low-resource ASR.
\end{abstract}

\section{Introduction}

Automatic speech recognition (ASR) systems have seen remarkable improvements in recent years, largely driven by the availability of large-scale labeled speech datasets and advances in deep learning. State-of-the-art models like Whisper\cite{radford2023robust} are trained on more than half a million hours of webscale transcripts. However, in many domains, acquiring high-quality transcriptions remains a costly and time-consuming process, limiting the amount of labeled data available for training. In fact, high-quality transcriptions can cost upwards of \$150 per hour of speech \cite{williams2011crowd, novotney2010cheap}. This high cost results in data scarcity in many domains which hinders the development of robust ASR models for low-resource and domain-specific tasks. 

A potential solution is to leverage weak transcriptions, which may be imperfect or noisy with many errors but still retain useful information and can be acquired cheaply. The central question we explore in this work is how to effectively integrate weak transcriptions with limited gold-standard transcriptions to improve ASR performance in low-resource domains. This approach falls under the domain of Weakly Supervised Learning (WSL) \cite{zhou2018brief}. Unlike fully supervised learning, where the model is trained on a precisely labeled dataset, and self-supervised or unsupervised learning where the model is trained on unlabeled data, WSL is a paradigm of machine learning where models are trained on partially labeled or imprecisely labeled data. 
% More precisely, our approach falls under the domain of \textbf{``inaccurate supervision"} where labels are not always precise.
% There are three types of WSL \cite{zhou2018brief}:
% \begin{itemize}
%     \item \textbf{Incomplete Supervision} where only a subset of the training data is labeled. Examples include self-training, where an intermediate model is first trained on the labeled subset, and then used to generate pseudo-labels for the available unlabeled data for subsequent training \cite{AMINI2025128904}. 
%     \item \textbf{Inexact Supervision} where only coarse-grained labels are given. For example, in image categorization tasks, where image-level labels instead of object-level labels are used.
%     \item \textbf{Inaccurate Supervision} where labels are not always precise.
% \end{itemize}

In the task of ASR, WSL, particularly inaccurate supervision \cite{frenay2013classification,brabham2008crowdsourcing}, has enabled the use of imprecise or inaccurate transcription. WSL has proved to be crucial in many ASR models, due to the large costs of accurate transcription. Perhaps the most well-known example of a weakly-supervised ASR model is Whisper, where the large training corpus was not precisely vetted, and portions of it were known to be inaccurate. However, the model's contextual capacity and the large size of the dataset proved to be adequate, so that the model can learn well from the noisy labels. 
% Perhaps the biggest study on weak transcriptions in ASR is Whisper itself. Whisper was primarily trained on weak web-scale transcriptions scoured from the internet. Although not a lot is known about Whisper's training data, we know that it also contained a substantial amount of off-the-shelf high-quality transcriptions and that the weak web-scaled transcripts were filtered in a way to remove the most inaccurate transcriptions. However, these weak transcriptions still affected Whisper's accuracy and reliability, as Whipser suffers from a well-documented issue with hallucinations, which are erroneous nonsensical transcriptions not related to the input audio \cite{ji2023survey}. A recent study \cite{baranski2025investigation}investigating the Whisper's hallucinations induced by non-speech audio found that among the most common hallucinations are phrases like ''subtitles by the amara org community" and ''thanks for watching", which reveals the effect of Whisper being trained on video transcriptions from movies or TV shows. However, overall, Whisper remains a very strong ASR system primarily due to its massive training data size which allowed it to outperform every other ASR model on average in general-purpose ASR. 

Whisper's success with inaccurate supervision motivated us to explore the applicability of weak transcripts in low-resource ASR, primarily in classroom speech. Classrooms present a unique challenge in ASR since they are multi-speaker environments with multiple target speakers (teachers and students) whose speech sometimes overlap \cite{chang2019end, chang2020end}. Additionally, classrooms are noisy environments characterized by children's babble noise, which is the noise of multiple background speakers and is considered one of the most difficult noises in ASR \cite{simic2024self}. However, there is a huge potential for sufficiently accurate ASR to help in the objective analysis and understanding of classroom dynamics\cite{demszky2023improving,jacobs2022promoting,jacobs2024automated}.

All of these challenges are complicated by the fact that classroom speech is a low-resource task \cite{attia2023kid, fan2024benchmarking} due to the protections around children \cite{coppa2024,cao2023comparative}. This means great care must be taken to best utilize existing data. One of the largest resources of classroom audio data is the National Center for Teacher Effectiveness (NCTE) classroom speech corpus \cite{demszky2022ncte}. The NCTE dataset contains 5000 hours of transcribed classroom speech, but these transcripts are very inaccurate as they were not intended for ASR training. They are not verbatim and all names of the students and teachers were de-identified to protect their privacy. These transcripts have not been used, and a small number of recordings were accurately re-transcribed using human gold-standard transcriptions \cite{attia2024cpt}. While Whisper has shown that inaccurate transcripts are useful if paired with sufficient accurate transcripts, it is not clear if imprecise transcripts in the NCTE dataset would be helpful, given that they far outnumber precise transcripts.

In this study, we explore how to best utilize these weak transcriptions to improve the performance of classroom ASR systems. We preface our experiments using the NCTE corpus with controlled experiments on synthetically corrupted transcripts from the TEDLIUM-3 \cite{hernandez2018ted} corpus.  

We propose Weakly Supervised Pretraining (WSP), a weakly supervised learning paradigm where we initially use imprecise transcription as an intermediate supervised step, followed by fine-tuning on a limited amount of gold-standard data. Our results indicate that this WSP paradigm is very effective in increasing the utility of limited gold-standard data, outperforming previous methods with limited classroom data.

% Our contributions in this paper are as follows:
% \begin{itemize}
%     \item 
% \end{itemize}
\vspace{-10pt}
\section{Datasets}
\subsection{TEDLIUM}
We use the TEDLIUM 3 \cite{hernandez2018ted} dataset for our synthetic experiments. TEDLIUM is a collection of transcribed recordings of TED talks amounting to 540 hours from 2028 speakers. We chose TEDLIUM as it is a high-quality and public dataset that provides a good benchmark for our experiments. Unlike Librispeech \cite{panayotov2015librispeech}, TED talks are a better representation of everyday spontaneous speech, making them more suitable for experiments with transcription errors.
\subsection{NCTE}
The NCTE dataset consists of video and audio recordings of 2128 4th and 5th-grade elementary math classrooms \cite{demszky2022ncte}. Each classroom is recorded with 2 to 3 microphones that range from lanyards worn by teachers capturing near-field speech to far-field stationary microphones placed in the corner of the classroom. In total, the dataset contains 5235 hours of recordings. 

\subsubsection{NCTE-Weak}
The majority of these recordings were transcribed, but they were not intended for ASR tasks. These transcriptions contain a large number of substitutions and deletions, and all the names of students and teachers were omitted to protect their privacy. Most importantly, the transcripts were not properly time-stamped. Loose inaccurate minute marks are provided that are often off by tens of seconds. In addition, in many instances, utterances have been labeled as ``unintelligible" or ``side conversation" that are still perfectly intelligible to both ASR models and humans. Thus far, these transcriptions have not been used for ASR training, and previous works only utilized them for analysis \cite{demszky2022ncte} or for training n-gram language models (LMs) for beam-search decoding \cite{attia2024cpt}. However, in this research, we attempt to utilize these transcriptions as weak transcriptions in intermediate training. We call this dataset \textbf{NCTE-Weak}. We pass the dataset through a forced aligner \cite{MahmoudAshraf_mms300m1130_aligner, wolf2019huggingface} to obtain more accurate timestamps for preprocessing the dataset before ASR training. While the forced aligner gave better timestamps than the ones provided, deletion errors and the mismatch between the forced alignment model and the classroom speech domains resulted in errors in timestamps. 
\vspace{-5pt}
\begin{tcolorbox}[colback=gray!10, colframe=gray!50, arc=2mm, boxrule=0.5pt, left=0pt, right=0pt, top=0pt, bottom=0pt]
\scriptsize
    \textbf{Ground Truth:} \textit{So what would your final answer be?}\\
    \textbf{Provided Transcription:} \textit{line, he is using a number line. Good. So then what would your final answer be?}
\end{tcolorbox}
\vspace{-5pt}

 % \end{quote}

\subsubsection{NCTE-Gold}
In previous works\cite{attia2024cpt}, a small subset of the recordings were re-transcribed using gold-standard human transcriptions. In this work, we transcribe 11 more classrooms raising the number of gold-standard transcribed classroom recordings to 17 classes, amounting to 13 hours of recordings. We call this subset \textbf{NCTE-Gold} and we split it into a training set (10 hours from 13 classes) and a validation set (3 hours from 4 classes).

\section{Experiments}
In this section, we outline our experiments to better understand the effect of weak transcriptions on ASR performance and showcase the efficacy of WSP. First, we perform an ablation study using the TEDLIUM dataset where we synthetically corrupt increasing portions of the training data with common transcription mistakes. We then apply our findings to real-world scenarios using the NCTE dataset. For both experiments, we train Wav2vec2.0-based models, using the fairseq \cite{wang2020fairseq} implementation. We chose to use Wav2vec in our experiments as it does not have an internal LM like Whisper, which allows for better measurement of the effect of weak transcripts on ASR.

\subsection{Synthetic Corruption}
For the synthetic corruption experiments, we finetune Robust-wav2vec \cite{hsu2021robust}, as it is the Wav2vec model pre-trained with the largest amount of English speech audio.  

\subsubsection{Weakly Supervised Pre-training Step}
For the intermediate WSP step, we corrupt the transcriptions to model and approximate common human transcription mistakes. We consider 3 types:
\begin{itemize}
    \item \textbf{Deletion:} We randomly drop words from the sentence.
    \item \textbf{Misspellings:} We use the Datamuse API \cite{DatamuseAPI}  to replace a word with another word that either sounds like it to model hearing mistakes, or one that is spelled like it, to model common spelling mistakes.
    \item \textbf{Timestamp Inaccuracies:} Long recordings are often partitioned into smaller utterances; however, the timestamps for the beginning or end can be inaccurate. We model this inaccuracy by dropping a few words from the beginning and/or end of a sentence or adding a few random words. 
\end{itemize}
We use two methods to corrupt each sentence:
\begin{itemize}
    \item \textbf{Random Corruption:} Where at least one or more of the above mistakes are randomly applied, but the rest of the sentence is mainly correct. Each word had a 5\% probability of being deleted, a 20\% probability of being replaced by a soundalike, or a common misspelling, and a 5\% probability of being repeated. Each sentence had a 50\% probability of timestamp inaccuracies.
    \item \textbf{Full Corruption:} To model extreme corruption, every word in the sentence is misspelled or deleted. The rest followed the same paradigm as random corruption. %This was inspired by Lead2Gold's method \cite{dufraux2019lead2gold} of training a weak ASR model to generate weak transcriptions. However, our method is more methodical since it is driven by common transcription mistakes. 

\end{itemize}

Below is an example of random and full corruption as compared to the ground truth transcription. 

% \begin{quote}
\vspace{-5pt}
\begin{tcolorbox}[colback=gray!10, colframe=gray!50, arc=2mm, boxrule=0.5pt, left=0pt, right=0pt, top=0pt, bottom=0pt]
\scriptsize
    \textbf{Ground Truth:} \textit{Was offered a position as associate professor of medicine.}\\
    \textbf{Random Corruption:} \textit{Okay was offeree a position as associate of medicine.}\\
    \textbf{Full Corruption:} \textit{Coffered a exposition exposition ass assonate professore off medicines.}
    \end{tcolorbox}
\vspace{-5pt}
For each corruption method, we interpolate the corrupted transcriptions with gold-standard accurate transcriptions to create different training sets with different degrees of corruption. We create 4 training configurations for each corruption method at 25\%, 50\%, 75\% and 100\% corruption, where the percentages refer to the number of corrupted utterances in the training set. We also train the same model using full uncorrupted transcriptions as a baseline. We trained the models for 150,000 steps, and all models converged around 70,000 steps.
\subsubsection{Precise Fine-tuning}
All the models in the previous sections are then fine-tuned using 10 minutes of precisely transcribed and uncorrupted data. This approximates an extremely low-resource scenario.

\subsection{Real World Case Study - Classroom Data}
For a real-world case study, we consider the NCTE dataset. We use CPT-Boosted Wav2vec2.0\cite{attia2024cpt} which was specifically adapted to classroom speech using the NCTE dataset.
\subsubsection{Weakly Supervised Pre-training Step}
% We first train the intermediate model using the \textbf{NCTE-Weak}. We experimented with omitting sentences with names that are de-identifed altogether or leaving them in and considering the names to be deleted. We found that removing these utterances did not affect the final performance, indicating that the inaccuracies and other insertions, deletions, and substitutions have a more severe effect on training. We only list the results from the model with these utterances left in for space reasons. 

We first train the intermediate model using the \textbf{NCTE-Weak} dataset. We partition the dataset into training and validation splits with no test data using a 90/10 split after removing all the recordings that were re-transcribed in the NCTE-Gold dataset to prevent data leakage. We test that model on the test set from NCTE-Gold. We trained the model using the configuration file used for training Wav2vec on all 960 hours of Librispeech which trains the model for 320,000 steps.   
\subsubsection{Precise Fine-tuning}
We set the model trained on NCTE-Weak as initialization for our precise fine-tuning. We fine-tune that model using the NCTE-Gold training data for 500 steps using a few-shot learning scenario. We compare that model against two base-lines:
\begin{itemize}
    \item \textbf{Direct Fine-tuning: }we directly fine-tune the CPT-Wav2vec2.0 model for 20,000 steps on NCTE-Gold. \cite{attia2024cpt}
    \item \textbf{Self-training: \cite{grill2020bootstrap, AMINI2025128904}} We use an already fine-tuned model to transcribe a portion of the NCTE-Weak dataset and add it to the NCTE-Gold dataset. This adds 40 hours to the NCTE-Gold dataset. We fine-tune the CPT-Wav2vec2.0 model using this dataset for our second baseline.
\end{itemize}
\section{Results and Discussion}
\subsection{Synthetic Corruption}
% In this section, we discuss the results of our experiments training Wav2vec with corrupted TEDLIUM transcriptions. We show Word Error Rates (WERs) for each model with the original uncorrupted TEDLIUM test set. We consider the raw output of the model using greedy decoding as well as language model decoding. For the TEDLIUM experiments, we used the news text language model packaged with the dataset as it yielded the best performance.

We test each model regardless of corruption type or percentage on the uncorrupted TEDLIUM test set. We consider the results with and without LM beam-search decoding. Table \ref{tab:ted-corr} gives the Word Error Rate (WER) results for each model.

 The first row shows the performance of the model trained on the original dataset without any corruption. The results from the randomly corrupted training sets in the second column, show that while random inconsistent transcription mistakes negatively affect the model’s performance, they are not detrimental overall. We can see that even when half of the training data had transcription mistakes, the performance was barely affected and the degradation in WER was less than 1\%. Even when all the transcriptions in the training data had random mistakes, the performance was degraded by about 2\% with greedy decoding, and less than 1\% with LM decoding. 

In contrast, in fully corrupted training sets, where every word in the corrupted sentence was misspelled, the degradation in performance is noticeably higher. In the 25\% and 50\% corrupted training sets, the model could still perform reasonably well \textbf{with LM decoding}, however, the performance with non-LM greedy decoding is noticeably worse. This highlights one important distinction between the two corruption methods, where the gap between greedy decoding and LM beam search decoding is significantly larger with full corruption. This points to the fact that most of the misspelling mistakes are usually caught by the LM and corrected accordingly. While this doesn't work as well with higher degrees of corruption (75\% and 100\%), the improvement in performance with LM decoding is still significant,  although the utility of the models at this level of performance is limited.
\begin{table}[]
\centering
\caption{WER results from training Wav2vec2.0 with corrupted TEDLIUM transcriptions. “\% Corr.” indicates the proportion of utterances corrupted in the training set. Results before the slash (/) are from greedy decoding (no LM), while those after the slash use LM decoding.} \vspace{-5pt}
\label{tab:ted-corr}
\resizebox{0.65\columnwidth}{!}{%
\begin{tabular}{c|c|c}
\hline
\textbf{\% Corr.} & \textbf{Random Corr.} & \textbf{Full Corr.} \\\hline
\textbf{0} & \multicolumn{2}{c}{\textbf{7.40/6.77}} \\
\textbf{25} & 7.50/6.82 & 12.72/8.16 \\
\textbf{50} & 7.89/7.06 & 46.73/10.57 \\
\textbf{75} & 8.76/7.03 & 72.45/ 37.76 \\
\textbf{100} & 9.73/7.72 & 80.36/52.40\\\hline
\end{tabular}%
}\vspace{-10pt}

\end{table}

The results of fine-tuning the models trained on corrupted data on 10 minutes of accurate gold-standard data are in Table \ref{tab:ted-ft}. We also fine-tuned the model that was trained on the full original uncorrupted training data as a baseline, which is shown in row 1. We can see that fine-tuning with a small amount of labeled data seems to cancel out the effect of random inconsistent corruption in the original training data with less than 50\% random corruption. Further fine-tuning these models matches the performance of the model trained without any corruption in the first row of Table \ref{tab:ted-corr}. With higher degrees of corruption, we still see some improvement with greedy decoding, but we also see, interestingly, that with LM decoding, the model originally trained with some random corruption in the transcription in 100\% of the training labels slightly underperforms the model trained without any corruption.

With models that were originally trained with fully corrupted labels, fine-tuning with 10 minutes of accurate data helps them become more usable. With LM decoding, even with up to 50\% full corruption of the training labels, fine-tuning on a small amount of labeled data reduces the performance gap caused by corruption to below 2\%. Even with 75\% and 100\% corruption in the training data, fine-tuning on 10 minutes of accurate data reduces the WER significantly, by around 40\% with greedy decoding, and 25\% with LM decoding. 

\begin{table}[]
\centering
\caption{WER results from finetuning models from Table \ref{tab:ted-corr} on 10 minutes of accurate uncorrupted TEDLIUM data. “\% Corr.” indicates the proportion of utterances corrupted in the training set of the pretrained model.}\vspace{-5pt}
\label{tab:ted-ft}
\resizebox{0.65\columnwidth}{!}{%
\begin{tabular}{c|c|c}
\hline
\textbf{\% Corr.} & \textbf{Random Corr.} & \textbf{Full Corr.} \\\hline
\textbf{0} & \multicolumn{2}{c}{7.27/\textbf{6.48}} \\
\textbf{25} & \textbf{7.11}/6.50 & 10.31/7.53 \\
\textbf{50} & 7.49/6.65 & 13.30/8.33 \\
\textbf{75} & 8.10/6.55 & 33.91/19.0 \\
\textbf{100} & 8.31/6.86 & 43.90/ 24.49\\\hline
\end{tabular}%
}
\vspace{-21pt}
\end{table}

To understand the impact of such results, we note that correcting 10 minutes of weak transcription by a human can take less than an hour, as each minute of transcription can take between 5-8 minutes \cite{transcription_speed, cieri2004fisher} and can cost around \$25 \cite{williams2011crowd}. However, a model trained with 100\% of its training data fully corrupted, a very extreme case, can get a WER as low as 24.49\% with fine-tuning on just 10 minutes of accurate data. While 24.49\% is a high WER, it can be usable for many low-resource tasks and matches some baselines in classroom ASR as seen in Table \ref{tab:class}. 

Finally, we attempted to train the model directly using 10 minutes and 1 hour of labeled data without prior initial training, but the model did not converge. We did not include that model in the table, but the reader can assume its error to be 100\% since it did not learn to output any characters. While Wav2vec2.0 has been successfully trained with just 10 minutes of Librispeech before, Librispeech is a much cleaner and more straightforward task than TEDLIUM. This highlights an important finding: not only does further fine-tuning on accurate data cancel out the effect of moderate corruption in the transcription, but even severe corruption significantly improves the utility of small precise data. While it remains potentially possible to train a Wav2vec model using limited amounts of data from TEDLIUM, the fact remains that prior training on severely corrupted data makes this training much more straightforward. 
% While the 10 minutes and even 1 hour of data were not enough for the model to converge to any meaningful output, initial training on extremely corrupted and inaccurate transcription (100\% full corruption) paired with this small data resulted in meaningful and somewhat useful ASR models.

\subsection{Real World Case Study - Classroom Data}

\begin{table}[]
\centering
\caption{WER results on the NCTE and MPT test sets with different training configurations. NCTE-Weak $\rightarrow$ TED 10-Hr and NCTE-Weak $\rightarrow$ NCTE-Gold refer to models initially trained on the NCTE-Weak dataset and then finetuned on 10 hours of TEDLIUM and the NCTE-Gold dataset respectively. }
\label{tab:class}
\resizebox{0.9\columnwidth}{!}{%
\begin{tabular}{l|l|l}
\hline
\textbf{Training Data} & \textbf{NCTE} & \textbf{MPT} \\ \hline
\textbf{TEDLIUM} & 55.82/50.56 & 55.11/50.50 \\ 
\textbf{NCTE-Weak} & 36.23/32.30 & 50.84/46.09 \\
\textbf{NCTE-Gold} & 21.12/16.47 & 31.52/27.93 \\ 
\textbf{NCTE-Self Training} & 17.45/15.09 & 27.42/26.24 \\\hline
\textbf{NCTE-Weak → TED 10-Hr} & 25.59/21.14 & 42.62/37.22 \\
\textbf{NCTE-Weak → NCTE-Gold} & \textbf{16.54/13.51} & \textbf{25.07/23.70} \\ \hline
\end{tabular}%
}
\vspace{-20pt}
\end{table}

Looking at Table \ref{tab:class}, we see that the model trained on TEDLIUM performs poorly with both test sets which corroborates previous research \cite{attia2024cpt, southwell2024automatic} that shows that ASR models trained on off-the-shelf data struggle with classroom speech. While training on the NCTE-Weak improves the results, it is outperformed by training on NCTE-Gold, a precisely transcribed dataset that is 500 times smaller. Using the resultant model to generate self-training data further improves the results on both datasets. 

We ran two WSP experiments, both relying on first training on NCTE-Weak followed by precise fine-tuning. First, we consider the case where no in-domain accurate data exists. We use 10 hours of TEDLIUM, which we have already seen to be a poor match for classroom speech as evident from the first row of Table \ref{tab:class}. However, few-shot fine-tuning on this dataset improves the results significantly from NCTE-Weak training, by about 10\% in both configurations, while still underperforming direct precise fine-tuning on in-domain data. If in-domain accurate data exists, such as NCTE-Gold, we get further improvements, resulting in our best configuration for both NCTE and MPT test sets. Unlike our experiments with synthetic corruption, we do not have an estimate of the performance had we trained on 5000 hours of accurately transcribed NCTE recordings. Our previous results indicate that this configuration might approximate this performance. In any case, this configuration outperforms both direct fine-tuning and self-training, further showing that initial imprecise training increases the utility of limited precise data. 
\section{Error Analysis}
In this section, we discuss sample transcription from models trained with weak supervision, and those further fine-tuned on small accurate data. With this analysis, we attempt to understand the effect of transcription errors on performance, and why initial large-scale weak pre-training improves performance. 

\subsection{Synthetic Corruption}
The example below shows a sample transcription from the extreme case of 100\% of the training data being fully corrupted as the WSP model, and model that was subsequently fine-tuned on 10 minutes of precise data. WER is shown in parentheses. 
% \begin{quote}
\vspace{-5pt}
\begin{tcolorbox}[colback=gray!10, colframe=gray!50, arc=2mm, boxrule=0.5pt, left=0pt, right=0pt, top=0pt, bottom=0pt]
\scriptsize
    \textbf{Ground Truth:} \textit{Everybody talks about happiness these days.}\\
    \textbf{100\% Full Corr. (WSP):} \textit{e bod tal abou hapne thel da. }(116.67\%) \\
    \textbf{WSP $\rightarrow$ 10 min FT:} \textit{Ever body talks about hapines thees das.} (83.34\%)
% \end{quote}
\end{tcolorbox}
\vspace{-5pt}

At first glance, the transcription from the WSP model might seem complete gibberish. However, we can see that the model correctly detects some phonemic features from the sentence. For example ``bo"  matches ``everybody", ``ta" matches ``talks", etc. This can be attributed to the fact that even though the majority of words in the training set were replaced by their soundalikes or common misspellings, these alternative words still match a lot of the sounds in the audio, which the model partially succeeds in detecting.  Further precise fine-tuning builds on this weak foundation to output a legible transcription with a few spelling and structure mistakes. These mistakes are usually fixed by LM beam search decoding. In fact, LM decoding reduces the WER for this example to 16.67\% with a single substitution error (day $\rightarrow$ das).

\subsection{Real World Case Study - Classroom Data}
The example below shows a sample transcription from three models, the model directly trained on NCTE-Gold, the model trained on NCTE-Weak (WSP), and the model initially trained on NCTE-Weak and then fine-tuned on NCTE-Gold. The models do not predict casing or punctuation, however they were added to improve legibility.\\
% \begin{quote}
\vspace{-15pt}
\begin{tcolorbox}[colback=gray!10, colframe=gray!50, arc=2mm, boxrule=0.5pt, left=0pt, right=0pt, top=0pt, bottom=0pt]
\scriptsize
    \textbf{Reference:} \textit{33. You are right. Yours is correct okay. I won. Clean up. Go back to your seat. Do not put anything away. You are going to need it. 54. Faithful, you have to stay there, you can not move because of the camera.}\\
    \textbf{NCTE-Weak (WSP):}  \textit{33. You re right. Yours is correct. 54. You have to stay there. n  me csthe me.} (75.00\%) \\
    \textbf{NCTE-Gold:} \textit{33. You are right. Yours is correct. kay. on one. Cleen up. Go back to your sseat. Do not put anything away, you are going to need it. 54. Faithfuel, you have to stay there, you can not move because of the camera.} (15.91\%) \\
    \textbf{WSP $\rightarrow$ NCTE-Gold:} \textit{33. You are right. Yours is correctokay. On one. Clean up. Go back to your seat. Do not put anything away, you are going to need it. 54. Faithful, you have to stay there, you can not move because the camera. }(11.36\%)
\end{tcolorbox}

% % \end{quote}
We can that the worst performance is when we train with NCTE-Weak. Similarly to how the training data suffered from deletion errors, the transcription skips over large portions of the audio and picks up later. For instance, the phrases ''\textit{Clean up ... You are going to need it.}" are completely missing, followed by correctly transcribing the number 54, and so on. By inspecting the audio, there does not seem to be any pattern to the deleted portions in tone or noise level. However, the model picks up the initial phrases of the audio correctly. The model often gets 0\% error for shorter instances. This might highlight some ways the utility of this dataset can be improved further. 

However, even with the deletion mistakes, the model often captures portions of the audio correctly, which explains why fine-tuning this model with NCTE-Gold outperforms direct fine-tuning on the same dataset. By comparing both transcriptions, we see that initial WSP improves several aspects of the transcription, such as the misspelling on ''\textit{clean}"  as ''\textit{cleen}", and the substitution of the name Faithful.  Fine-tuning on NCTE-Gold does not only fix the deletion mistakes caused by timestamp inaccuracies in NCTE-Weak, but also utilizes the exposure to large amounts of weak data, resulting in improved robustness.
\section{Conclusion}
In this paper, we introduced Weakly Supervised Pretraining (WSP) as an effective approach for leveraging highly inaccurate transcriptions in low-resource ASR settings. Our synthetic corruption study on TEDLIUM demonstrated that models trained on even severely corrupted transcriptions, when followed by fine-tuning on just 10 minutes of precise data, can achieve usable ASR performance. We extended this insight to real-world classroom speech using the NCTE dataset, showing that WSP on weak transcripts outperforms direct fine-tuning on limited accurate data and even outperforms self-training approaches.

\bibliographystyle{IEEEtran}

\bibliography{mybib}

\end{document}